\documentclass[reprint, amsmath,amssymb,]{revtex4-2}
\usepackage{graphicx}
\usepackage{dcolumn}
\usepackage{bm}
\usepackage{hyperref}
\usepackage{color}
\usepackage{siunitx}
\usepackage{mathrsfs}
\usepackage{amsmath}
\usepackage{textcomp}
\usepackage{natbib}
\usepackage[utf8]{inputenc}
\makeatletter
\newcommand*\bigcdot{\mathpalette\bigcdot@{.7}}
\newcommand*\bigcdot@[2]{\mathbin{\vcenter{\hbox{\scalebox{#2}{$\m@th#1\bullet$}}}}}
\makeatother
\makeatletter
\usepackage[explicit]{titlesec}

\titlespacing*{\section}{0pt}{4.0ex plus .8ex minus 0.5ex}{1.6ex plus .0ex}
\titlespacing*{\subsection}{0pt}{3.5ex plus .0ex minus .0ex}{2.3ex plus .0ex}
\hypersetup{hypertex=true, colorlinks=true, linkcolor=blue, anchorcolor=blue, citecolor=blue}

\begin{document}
	
	\preprint{APS/123-QED}
	
	\title{Single-Spin Waved-Brim Flat-Top Hat in the Band Edge of GdIH Monolayer}

	\author{Ningning Jia\textsuperscript{1}}
	
	\author{Zhao Yang\textsuperscript{1}}
	
	\author{Jiangtao Cai\textsuperscript{2}}

	\author{Zhiheng Lv\textsuperscript{1}, Yongting Shi\textsuperscript{1}, Tielei Song\textsuperscript{1}, Xin Cui\textsuperscript{1}
	}

	\author{Zhifeng Liu\textsuperscript{1}}
	\email{Corresponding author: zfliu@imu.edu.cn(Z.Liu)}

	\affiliation{%
		$^1$Key Laboratory of Nanoscience and Nanotechnology, School of Physical Science and Technology, Inner Mongolia
		University, Hohhot 010021, China\\
		$^2$School of Physics and Information Science, Shaanxi University of Science and Technology, Xi’an 710021, China}
	
	\date{\today}
	
	\begin{abstract}
		\noindent Exotic electronic bands, such as flat bands, linear crossing bands, spontaneously valley- or spin-polarized bands, in two-dimensional materials have been the hot topics in condensed matter physics. Herein, we first propose a general dispersion model for possible hat-like electronic bands, and then identify an intriguing single-spin  \emph{waved-brim flat-top hat} in the valence band edge of a stable ferromagnetic semiconducting electrene (i.e., Janus GdIH monolayer), which can be well described by a simplified two-bands Hamiltonian model. Specifically, the hat-band has a waved brim with six valleys along the boundary of the first Brillouin zone; meanwhile it holds a flat top close to the Fermi level, resulting in the emergence of single-spin van Hove singularities divergence and Lifshitz transitions. Owing to the breaking of both time-reversal and space inversion symmetries, a sizable spontaneous valley polarization is formed between the adjacent brim valleys, which provides the opportunity to realize the high-temperature anomalous valley Hall effect. Particularly, via modest strains and carriers doping, various conductive bipolar-states (spin-up vs. spin-down, K valley vs. $-$K valley, and ultra-low-speed vs. ultra-high-speed) can be modulated out from the distorted waved-brim flat-top hat of GdIH ML.  
	\end{abstract}
	\maketitle
	\UseRawInputEncoding
	
	\section{INTRODUCTION}
    \hspace{1.0em}In condensed matter physics, the electronic behaviors of solid-state materials have always been a central topic \cite{RN1}. The band theory tells us that electronic properties of a crystal mainly depend on its band structure \cite{RN2,RN3}, especially its valence or conductive edges in the vicinity of Fermi energy. Over the past decades, many exotic frontier bands, such as flat bands \cite{RN4}, linear Dirac/Weyl bands \cite{RN5, RN6}, half-metallic bands \cite{RN7}, and ferrovalley bands \cite{RN8}, have been identified with fantastic features and promising applications. In a sense, the discovery of a new distinguished band often triggers new physics, applications, and even manufacturing chains. 
	
	As is known, Bloch electrons in periodic crystals have three intrinsic degrees of freedom: charge, spin, and valley (the local minimum or maximum of band edges). The utilization of them places special requirements on the electronic bands, and then gives birth to different subdisciplines, corresponding to microelectronics \cite{RN9}, spintronics \cite{RN10}, and valleytronics \cite{RN11}. As a major factor, the band dispersion determines the movement behavior of charge carriers in crystal materials. Generally, the effective mass \( m^* \), defined as ${\hbar^2}/{m^*} ={\rm d}^2E(\textbf{\textit{k}})/{{\rm d}\textbf{\textit{k}}^2} $, is a key parameter to reflect band dispersion and can take three cases: \( m^* = {\rm C} \)(a finite constant), \( m^* \to 0 \), and \( m^* \to \infty \). For the latter two extreme ones, the local band dispersion should be linear and flat, respectively, which have attracted numerous attentions in past few decades \cite{RN4, RN5, RN6}. Up to date, the linear Dirac/Weyl dispersions have been widely identified in topological insulators \cite{RN12} and topological semimetals \cite{RN5}; additionally, dispersion-less flat/quasi-flat states have been detected in Kagome lattice \cite{RN13}, moir\'{e} superlattice \cite{RN14}, and many other lattices (e.g., pyrochlore, Lieb, bipartite) \cite{RN4}.

	Since the spins of electrons can be used as information carriers, the spin transport electronics  is considered as a promising methodology for increasing the efficiency of data storage and transfer in next generation information technology \cite{RN10}. For applications, a prerequisite is to generate or inject spin in the real materials, which is one of the major challenges in spintronics \cite{RN6}. In this context, the intrinsic ferromagnetic materials with high or even 100$\%$ spin-polarization are highly desirable. In solid bands, the significant lifting of spin degeneracy can usually be induced by magnetic exchange interaction. Based on different feature of band dispersion, various fully spin-polarized bands have been proposed so as to meet different spin-transport needs and physical applications, such as Half-Dirac bands \cite{RN15}, spin gapless semiconducting bands\cite{RN16}, single-spin flat bands \cite{RN17}, and bipolar magnetic semiconducting bands \cite{RN18}.
	
	As an analogue of spin, valley (termed as pseudospin) can also be employed to store and carry information due to that a large separation exists between inequivalent valleys in momentum space and intervalley scattering is strongly suppressed \cite{RN19}. In valleytronics like that in spintronics, there is a big challenge to generate valley polarization. Although some external strategies (e.g., optical pumping \cite{RN20}, magnetic doping \cite{RN21}, magnetic proximity \cite{RN22}, external magnetic field \cite{RN23} and electric field \cite{RN24}) have been experimentally confirmed to be effective for eliminating the valley degeneracies, many insurmountable drawbacks (e.g., limited carrier lifetimes, clustering effect, enlarged device size, and weak valley polarization)  still exist \cite{RN25}, which has greatly hampered the development of valleytronics. For this, the ferrovalley band with intrinsic spontaneous valley polarization was creatively proposed by Tong et al. \cite{RN8} in 2016, which may open a new avenue for valleytronics. When the extreme band dispersion is further considered, the interesting half-valley metal \cite{RN26} or valley-half-semimetal \cite{RN27} states with linear Dirac bands have also been established for realizing valley-dependent ultra-high-speed quantum transport. 
	
	In the mentioned exotic bands, people usually just focus on a certain aspect related to the degrees of freedom of Bloch electrons. For the miniaturization of next generation electronic devices, however, it is mush needed to integrate and manipulate multiple functions in only one nanomaterial. Therefore, it would be exciting to realize the combination and flexible adjustability of charge, spin, valley and even extreme dispersion in just one special band. In this work, we establish a general dispersion model for possible hat-like bands, and subsequently demonstrate that a new kind of hat-like dispersion, termed as \textit{waved-brim flat-top hat} which can be derived from a simple two-bands Hamiltonian model, can meet the above expectations. Using first-principles calculations, we show that the ferromagnetic 2D electrene, Janus GdIH monolayer (ML), harbors a single-spin waved-brim flat-top¬ hat in its valence band edge. Moreover, the structure, stability, magnetic properties of GdIH ML have been systematically investigated. In particular, the possible new physics and potential applications about the single-spin waved-brim flat-top¬ hat band have also been discussed.

	\section{COMPUTATIONAL METHODS}
	
	\hspace{1.0em}In our first-principles calculations, all the geometric optimizations and electronic structure calculations are performed in the framework of density functional theory (DFT) by using the Vienna \textit{ab initio} simulation package (VASP) \cite{RN28}. The generalized gradient approximation (GGA) of the Perdew-Burke-Ernzerhof (PBE) functional is adopted to deal with the change-correlation interactions \cite{RN29}. A plane-wave basis with a cutoff energy of 600 eV is employed to expand the wave functions. A Monkhorst-Pack \textit{\textbf{k}}-point mesh \cite{RN30} with a uniform spacing of $2{\pi}{\times}$0.02 {\AA}$^{-1}$ is used to sample the Brillouin zone (BZ). During the structural relaxation, the convergence criteria of total energy and residual Hellmann-Feynman force on each atom are set to $10^{-6}$ eV and  0.001 ${\rm eV}/${\AA}$^{-1}$, respectively. To avoid the interaction between the adjacent periodic images, a large vacuum space of 20 {\AA} is applied along the \textit{z}-direction.  Considering the correlation effect, we adopt the GGA + U method with U = 4.6 eV for 4f of Gd, which has also been used in other Gd related systems \cite{RN31}. Furthermore, the calculation by Heyd-Scuseria-Ernzerh of hybrid functional method (HSE06) \cite{RN32} has also been performed to check the reliability of the electronic band structure of GGA + U. The Berry curvature is calculated by VASPBERRY code \cite{RN33} using ${\rm Fukui's}$ method \cite{RN34}. 
	\\\hspace*{\fill} \\                                                             
	\section{RESULTS AND DISCUSSION}
	
	\subsection{A general dispersion model for different hat-like bands.}
	\hspace{1.0em}In the last decade or so,  a curious band dispersion, shaped like a Mexican-hat \cite{RN35}, has been found in some 2D materials (e.g., bilayer graphene \cite{RN36,RN37}, III-V ${\rm Ga_2}$${\rm X_2}$ ML \cite{RN38}, penta-graphene \cite{RN39}), and even been experimentally confirmed in GaSe ML \cite{RN40}. Owing to the existence of 1D van Hove singularities (VHS) in the density of states (DOS) \cite{RN41}, such kind of 2D materials are proved to be promising in multiferroics \cite{RN41}, magnetism \cite{RN42}, thermoeletronics \cite{RN43}, and scaled transistors \cite{RN44}.

	Inspired by the reported Mexican-hat bands \cite{RN35,RN36,RN37,RN38,RN39,RN40,RN41}, here we first propose a general dispersion model based on 3D surface geometrics, which should be universal for describing the possible hat-like bands in the 2D systems. Its specific expression can be written as a trigonometric polynomial:
	\begin{equation}\label{func_1}
		E(k_x, k_y) = \sum_{j=0}^{m} \sum_{i=0}^{j} {\rm A}_{j-i,i} \cos[l(k_x - k_{x0})] \cos[i(k_y - k_{y0})]
	\end{equation}
	Thereinto, \textit{l} = $\textit{j} - \textit{i}$, the ${\rm A}_{j-i,i}$ are the adjustable parameters for different eigenvalues surfaces in a $k_x$-$k_y$ plane with \textbf{\textit{k}}-point $(k_{x0}, k_{y0})$  at the center; $m$ denotes as a positive integer, being no less than the natural numbers $i$ and $j$, namely $m \geq j \geq i$. For a finite value of $m$, the number of terms in the trigonometric polynomial $E(k_x, k_y)$ should be $N_{m+1} = N_m + (m + 2)$. The firm term is $N_1 = 3$ when $m = 1$. In principle, the larger the $m$, the more complex the energy surface which could be described. As the $m$ can be infinity, one can also view the function $E(k_x, k_y)$ as a series related to $m$.  
	
	\begin{figure*}[t]
		\includegraphics[width=0.65\linewidth]{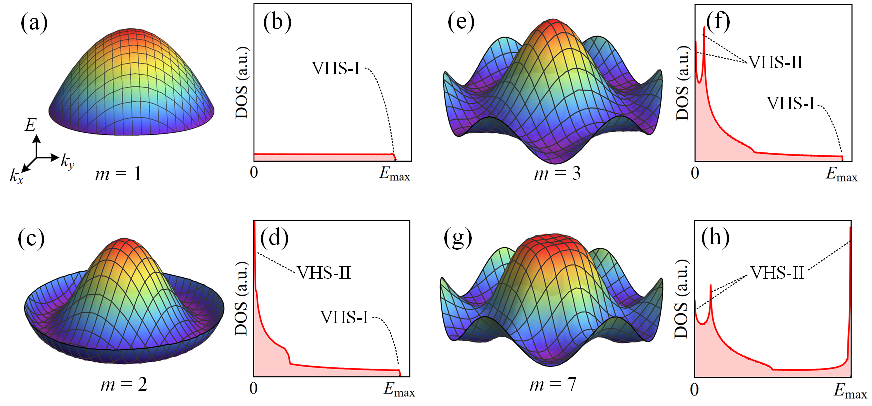}
		\caption{\label{fig:1}Four types of hat-like bands together with the corresponding DOS: (a-b) S-hat, (c-d) M-hat, (e-f) W-hat and (g-h) WF-hat.}
	\end{figure*}
	
	Taking \( (k_{x0}, k_{y0}) = (0, 0) \), we obtain four types of hat-like bands (see Fig. \ref{fig:1}) derived from the dispersion model Eq.1 with different \( m \) and parameters \( {\rm A}_{j-i,i} \):
	
	(1) \textit{Safety hat} (S-hat) band. When \( m = 1 \), one can easily deduce that \( E(k_x, k_y) = {\rm A}_{0,0} + {\rm A}_{1,0} \cos(k_x) + {\rm A}_{0,1} \cos(k_y) \). Then, a S-hat band without brim can be tuned out from the three parameters in the fractional Cartesian coordinate system of reciprocal space. Fig. \ref{fig:1}(a) displays the 3D \( E(k_x, k_y) \) surface in the area of \( k_x^2 + k_y^2 \leq 0.25 \). Such isotropic massive band dispersion (\( m^* \rightarrow {\rm C} \)) is quite common around high-symmetry \textit{\textbf{k}} points in the conductive band edge of 2D semiconducting materials.
	
	(2) \textit{Mexican hat} (M-hat) band. For \( m = 2 \), we can get a trigonometric polynomial with six terms:$E(k_x, k_y) = {\rm A}_{0,0} + {\rm A}_{0,1} \cos(k_y) + {\rm A}_{1,0} \cos(k_x) + {\rm A}_{0,2} \cos(2k_y) + {\rm A}_{1,1} \cos(k_x) \cos(k_y) + {\rm A}_{2,0} \cos(2k_x)$. By regulating the corresponding coefficients, the mentioned M-hat \cite{RN35,RN36,RN37,RN38,RN39,RN40,RN41} can be obtained [Fig. \ref{fig:1}(c)], which is consistent with the result of \( E(k) = E_0 + {\rm A}(k_x^2 + k_y^2)^2 + {\rm B}(k_x^2 + k_y^2) \) for \( {\rm A} > 0 \) and \( {\rm B} < 0 \) \cite{RN41}. Compared with the S-hat, M-hat has an extra tire-like brim. For this, the low energies Fermions around the brim can behave as the Fermi liquid due to the existence of the topological nontrivial Fermi surface, namely ``Fermi ring'' \cite{RN36}.
	
	(3) \textit{Waved-brim hat} (W-hat) band. When \( m \geq 3 \), one can modulate out an intriguing unknown W-hat dispersion using the different \( {\rm A}_{j-i,i} \) parameters. Taking the hexagonal systems as an example, a W-hat with \( {\rm C}_6 \) rotational symmetry can be obtained with \( m = 3 \), as is illustrated in Fig. \ref{fig:1}(e). Clearly, the body of the hat is similar with that of S-hat and M-hat, while the brim becomes more complex. One can see that there are six pairs of alternating extrema along the brim. Such local minimum or maximum of band edges could provide the opportunity for the utilization of the valley degree of freedom.  In this regard, the novel brim of W-hat should be promising for valleytronics \cite{RN11}. Next, we will present an in-depth discussion in conjunction with a specific material.
	
	(4) \textit{Waved-brim flat-top hat} (WF-hat). If \( m \) is up to 7, a more complex hat, i.e., WF-hat can be obtained by adopting 36 different \( {\rm A}_{j-i,i} \) coefficients. Compared with W-hat, the WF-hat inherits the similar waved-brim. However, it has an interesting quasi-flat top with almost negligible band dispersion [Fig. \ref{fig:1}(g)]. This means that the heavy Fermions (\( m^* \rightarrow \infty \)) exist in a very narrow energy window. Assuming that the WF-hat also has high mobility carriers in the brim valleys, the light and heavy Fermions may coexist in one material. This gives the possibility to modulate rich electronic states by coupling the velocity of carriers, spin, and valleys.
	
	On the basis of the obtained hat bands, the corresponding DOS can be calculated by
	\begin{equation}\label{func_2}
		g(E) = \frac{dn(E)}{dE} = \frac{1}{{(2\pi)}^2 \nabla_\textbf{\textit{k}} E} \frac{\rm dA_E}{{\rm d}\textbf{\textit{k}}} 
	\end{equation}in which, \( n(E) \) denotes the particle density at the energy \( E \), \( {\rm A}_E = \iint_{{\mathbb{\rm S}_{\rm E}}}\!\!\!\!\!\!\!\!\!\!\!\!\!\!\!\!\!\;\subset\!\supset  d\textbf{s} \) represents the occupied area of a specific energy surface \( {\rm S}_E \), and \( \nabla_\textbf{\textit{k}} E \) is the gradient of eigenvalues along possible \textbf{\textit{k}} directions. For a 2D system, the area that each state takes up in the momentum space is \( (2\pi)^2/{\rm V} \). Thus, we can get the particle density \( n(E) = {\rm A}_E/(2\pi)^2 \). On the basis of \( E(k_x, k_y) \), the DOSs of different hats are evaluated by discrete numerical calculations (see the right side of the corresponding 3D band in Fig. \ref{fig:1}). As marked in the pictures, these hat bands generate two types of VHSs: (i) VHS-I, a Heaviside step function discontinuity \cite{RN41} at maximum \( E = E_{\text{max}} \), (ii) VHS-II, a divergence in the DOS due to the appearance of saddle points with \( \nabla_\textbf{\textit{k}} E \rightarrow 0 \). Specifically, the S-hat has only VHS-I, while the WF-hat just holds VHS-II. The M-hat and W-hat possess both of VHS-I and VHS-II. Since the VHS-II is a strongly localized state, the materials with VHS-II are usually considered as an ideal platform for realizing interesting strongly correlated physics \cite{RN45,RN46,RN47}.
	
	From the above results, one can see that the WF-hat is more intriguing with respect to the other hats due to the coexistence of a flat-band state and brim valleys. To further exhibit the possible physics and practical applications, the matrix materials of WF-hat should be a central topic in the following researches. As spontaneous valley polarizations \cite{RN8} can be induced by breaking both time reversal and spatial inversion symmetry, hereafter we will focus on the single-spin WF-hat band in a new 2D FM material, i.e., the Janus GdIH ML. Before the specific discussions on the single-spin WF-hat, the structure and basic properties of the GdIH ML are firstly examined.
	
	\subsection{Structure, Stability, and Magnetic State of Janus GdIH ML}
	
	\hspace{1.0em}As the synthesized bulk GdI$_2$ holds the 2H-MoS$_2$-type vdW layered structure and its ML has a smaller exfoliation energy (0.24 J/m$^2$ \cite{RN31}) with respect to that of graphene (0.37 J/m$^2$ \cite{RN48}), many efforts have been devoted to the family of GdX$_2$ (X = F, Cl, Br, and I) MLs, focusing on their magnetic and electronic properties \cite{RN31,RN49,RN50}. In this work, we will just focus on the Janus GdIH ML due to that: (i) it has a particular WF-hat band edge, while other GdX$_2$ MLs hold W-hat band edge \cite{RN31,RN50}; (ii) in experiment, the similar Janus structure MoSH ML has been synthesized by substituting the top layer S atoms of 2H-MoS$_2$ with H atoms using gentle H$_2$ plasma treatment \cite{RN51}.
	
	\begin{figure*}[t]
		\includegraphics[width=0.65\linewidth]{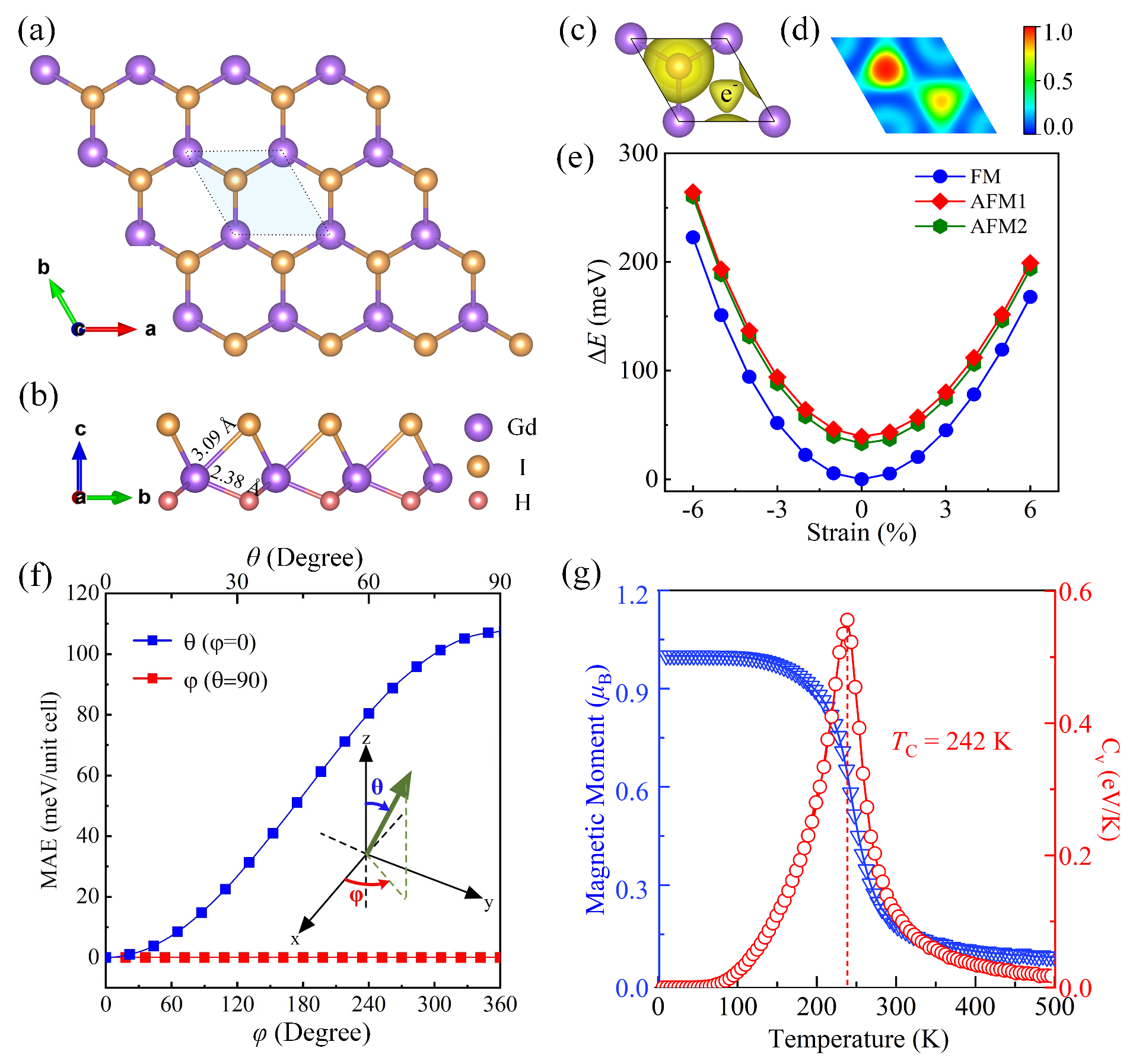}
		\caption{\label{fig:2}(a)Top and (b) side views for the optimized structure of GdIH ML. Its primitive cell is marked by the shaded area. (c) Electronic localization function (ELF) in real space and (d) its 2D map projected in the \textit{xy} plane. (e) Relative total energies of different magnetic states (FM, AFM1 and AFM2) as a function of external strains in the range from $-$6\% \text{ to } 6\%. (f) Magnetocrystalline anisotropy energy (MAE) in different magnetization directions of \textit{yz} plane (blue points) and \textit{xy} plane (red points). (g) The magnetic moment and heat capacity (${\rm C}_{\rm V}$) as a function of temperatures in the Monte Carlo simulation.}
	\end{figure*}
	
	\textit{Structure}. Like the MoSH ML \cite{RN51,RN52}, the optimized GdIH ML also adopts a closely packed sandwich structure [see Figs. \ref{fig:2}(a) and \ref{fig:2}(b)] and holds the trigonal space group symmetry (\(P3m1\), No. 156) with inversion asymmetry. Each intermediate Gd atom is bonded to three top-layer I atoms and three underlying H atoms, forming a trigonal prism coordination environment with \({\rm C}_{\rm 3v}\) point group symmetry. The lattice constants of GdIH ML are fully relaxed to \(a = b = 3.807 \ \text{\AA}\), being obviously smaller than that (4.099 \ \text{\AA} \cite{RN31}) of ${\rm GdI}_{2}$ ML due to the effect of hydrogenation. Another visual manifestation is that the bond length (3.09 \ \text{\AA}) of Gd-I is significantly longer than that (2.38 \ \text{\AA}) of Gd-H (see Fig. \ref{fig:2}(b)). The electron localization function \cite{RN53} [ELF, see Figs. \ref{fig:2}(c) and \ref{fig:2}(d)] of GdIH ML shows that the bonding of both Gd-I and Gd-H exhibits ionic feature. Additionally, one can find that there is a strongly localized electron (named as ``anionic'' electron \({\rm e}^-\)) at the center of a triangle Gd cluster, i.e., the interstitial area of the hexagonal lattice (Figs. \ref{fig:2}(c) and \ref{fig:2}(d)). This indicates that GdIH ML is a typical 2D electride material, i.e., electrene \cite{RN54}.  Considering the valence state of all atoms, the chemical formula of GdIH ML can be expressed as \({\rm Gd}^{3+}{\rm I}^-{\rm H}^- \cdot {\rm e}^-\), like the ideal ferrovalley semiconductor ${\rm LaH}_{2}$ ML \cite{RN55}.

	\textit{Stability}. To examine the stability of GdIH ML, we first evaluate its formation energy defined as: $E_{f} = (\mu_{\rm Gd} + \mu_{\rm H} + \mu_{\rm I} - E_{\rm GdIH})/3$, where $\mu_{\rm Gd}$, $\mu_{\rm H}$, and $\mu_{\rm I}$ are the chemical potentials of Gd, H, and I atom in their ground states, respectively. The obtained $E_{f}$ is 1.33 eV per atom, indicating that the formation of GdIH ML is an exothermic process. Since it is larger than that [1.23 eV, $E_{f}\acute{} = \left( \mu_{\text{Gd}} + 2\mu_{\text{I}} - E_{\text{GdI}_2} \right) / 3$
	] of $\rm GdI_{2}$ ML, the $\rm H_{2}$ plasma treatment like that in MoSH ML \cite{RN51} should be feasible for producing Janus GdIH ML. Then, we calculate the phonon spectrum of GdIH ML. As there no imaginary frequencies can be found (see Fig. S1 in the Supplemental Material), $\rm GdIH$ ML should be dynamically stable. The elastic constants are calculated to be ${\rm C}_{11} = 242.31$ N/m, ${\rm C}_{22} = 236.4$ N/m, ${\rm C}_{12} = 66.38$ N/m, and ${\rm C}_{44} = 82.17$ N/m, which can well satisfy the Born Huang criteria \cite{RN56}, including ${\rm C}_{11} > 0, {\rm C}_{11}{\rm C}_{22} > {{\rm C}_{12}}^{2}$, and ${\rm C}_{44} > 0$. Therefore, GdIH ML is also mechanically stable. For a 2D material, the gravity induced bending h/l can be estimated by the formula \cite{RN57}: $h/l \approx (\sigma gl/E_{\rm 2D})^{1/3}$,where $h$ is the out-of-plane deformation, $\sigma = 3.77 \times 10^{-6}$ kg/m$^2$ represents the surface density, the $l$ represents the size ($\sim$100 $\mu$m) of possible free-standing flake, the $g$ represents the gravitational acceleration, and $E_{\rm 2D}$ is the Young's modulus. With $\sigma = 3.77 \times 10^{-6}$ kg/m$^2$ and $E_{\rm 2D} = ({{\rm C}_{11}}^2 - {{\rm C}_{12}}^2)/{\rm C}_{11} = 224.13$ GPa$\cdot$nm, the $h/l$ of GdIH ML is evaluated to be $2.56 \times 10^{-4}$, which is of the same order of magnitude as that of graphene \cite{RN57} and ${\rm MnPSe}_{3}$ \cite{RN58} ML. This indicates that the stiffness of GdIH ML is sufficient to withstand its own weight like that of graphene and ${\rm MnPSe}_{\rm 3}$ ML.
	
	\textit{Magnetic ground state and its stability}. To determine the low-energy magnetic state of GdIH ML, we construct a redefined (2\textbf{a}, \textbf{a}+2\textbf{b}) supercell and consider the possible magnetic configurations, including paramagnetic (PM) state, ferromagnetic (FM) state, and different antiferromagnetic (AFM) states (see Fig. S2 in the Supplemental Material). Spin polarization is considered for FM and AFM states, but not for PM state. The calculated energies of these magnetic configurations show that the FM state is the most stable one, even under different mechanical strains [see Fig. \ref{fig:2}(e)], which are usually induced by the support substrate due to the existence of inevitable lattice mismatch \cite{RN59,RN60}. The optimized magnetic moment of FM GdIH ML is 1 $\mu_{\rm B}$ per unit cell. Owing to the feature of electrene, the magnetic moment is mainly contributed by the interstitial anionic electrons (see Fig. S2 in the Supplemental Material). In our calculations, the considered valence electron configurations of H, I, and Gd atoms are 1s$^1$, 5s$^2$5p$^5$, and 5p$^6$6s$^2$5d$^1$, respectively. Our Bader charge analysis \cite{RN61} shows that each H and I atom snatch one electron from the 6s state of the Gd atom, forming a stable closed-shell structures with anti-parallel spin, i.e., 1s$^2$ ($\downarrow\uparrow$) and 5p$^6$ ($\downarrow\uparrow\downarrow\uparrow\downarrow\uparrow$), respectively. Then the remaining 5d$^1$ of the Gd atom leaves its atomic central and forms an unpaired local anionic electron $e^-$ in the interstitial area due to the \textit{d}-\textit{d} hybridization among Gd atoms, giving rise to a magnetic moment of 1 $\mu_{\rm B}$.
	
	As the magnetocrystalline anisotropy (MCA) plays a critical role for stabilizing long-range FM ordering in a 2D material \cite{RN62,RN63}, it is necessary to examine the magnetic anisotropy energy (MAE) of GdIH ML, which often determines the capacity of FM ordering in resisting against heat fluctuation --- the larger the MAE, the better the thermal stability of the long-range FM ordering. By considering the spin-orbit coupling (SOC) effect, the relative energies along different magnetizing directions are obtained. From the Fig. \ref{fig:2}(f), one can see that (i) the easy axis of GdIH ML is along the out-of-plane (001) \textit{z} direction, like the synthesized $\rm CrI_{3}$ ML \cite{RN63} ; (ii) the magnetization in the \textit{x}-\textit{y} plane is isotropic, indicating that GdIH ML belongs to XY ferromagnet \cite{RN62}; (iii) the obtained MAE, the relative energy of the hard axis (100) or (010) with respect to the easy axis (001), is up to 107 \textmu eV, being significantly larger than the experimentally reported values of the Fe ML/Rh(111) (80 \textmu eV \cite{RN64}).
	
	To further explore the thermal stability of the FM ground state, we evaluate the Curie temperature ($T_{\rm C}$) by using different methods, including the mean-field approximation (MFA), Monte Carlo (MC) simulations, and first-principles molecular dynamics simulation. From the possible magnetic states (Fig. S2 in the Supplemental Material), the first ($J_{1}$) and second ($J_{2}$) nearest neighbor magnetic exchange interactions can be evaluated by using the classical Heisenberg model ${\rm H} = -\sum_{i,j} J_1 S_i S_j - \sum_{k,l} J_2 S_k S_l$.With the information of the coordination pattern of the local magnetic moments, the derived expression of $J_{1}$ and $J_{2}$ are $J_1 = -(E_{\text{AFM1}} - 2E_{\text{AFM2}} + E_{\text{FM}})/{8 m^2}$ and $J_2 = -(E_{\text{AFM1}} - E_{\text{AFM2}})/{4 m^2}$, where $E_{\text{FM}}$, $E_{\text{AFM1}}$, and $E_{\text{AFM2}}$ are the total energy of FM, AFM1, and AFM2, respectively, and $m$ denotes the local magnetic moment. For GdIH ML, the $J_{1}$ and $J_{2}$ are estimated to be 5.72 meV and 1.6 meV, respectively. Based on the obtained $J_{1}$ and $J_{2}$, the $T_{\rm C}$ is calculated to be 286 K by using MFA (see Text S1). Since the MFA method can't accurately describe the magnetic percolation effect \cite{RN65}, it usually overestimates the $T_{\rm C}$. For this, the ${T_{\text{C}}^{\rm MFA}}$ is further corrected by the empirical relationship $\left(T_{\rm C} / {T_{\text{C}}^{\rm MFA}}\right) = 0.51$ \cite{RN66}. Thus, a more reasonable $T_{\rm C}$ value should be $\sim$146 K. To examine this analytical result, we further perform numerical Monte Carlo (MC) simulation by taking the Wolff algorithm \cite{RN67} in a constructed $16 \times 16$ supercell. As is illustrated in Fig. \ref{fig:2}(g), the temperature dependent magnetic moment per unit cell and specific heat (${\rm C}_{\rm V}$) obtained from MC simulations reveal that the $T_{\rm C}$ of GdIH ML is about 240 K, which is well consistent with the MFA result and also our first-principle molecular simulation (see Fig. S1b in the Supplemental Material). Compared with the experimentally reported systems, e.g., CrI$_3$ (45 K \cite{RN63}), CrBr$_3$ (34 K \cite{RN68}), PTC-Fe (15 K \cite{RN69}), and Fe-T4PT (1.8 K \cite{RN70}), the proposed GdIH ML should be a more promising 2D magnet for spintronic applications in the higher temperature.
	
	\subsection{The WF-hat Band in Valence Band Edge of GdIH ML}
	
	\begin{figure*}[t]
		\includegraphics[width=0.85\linewidth]{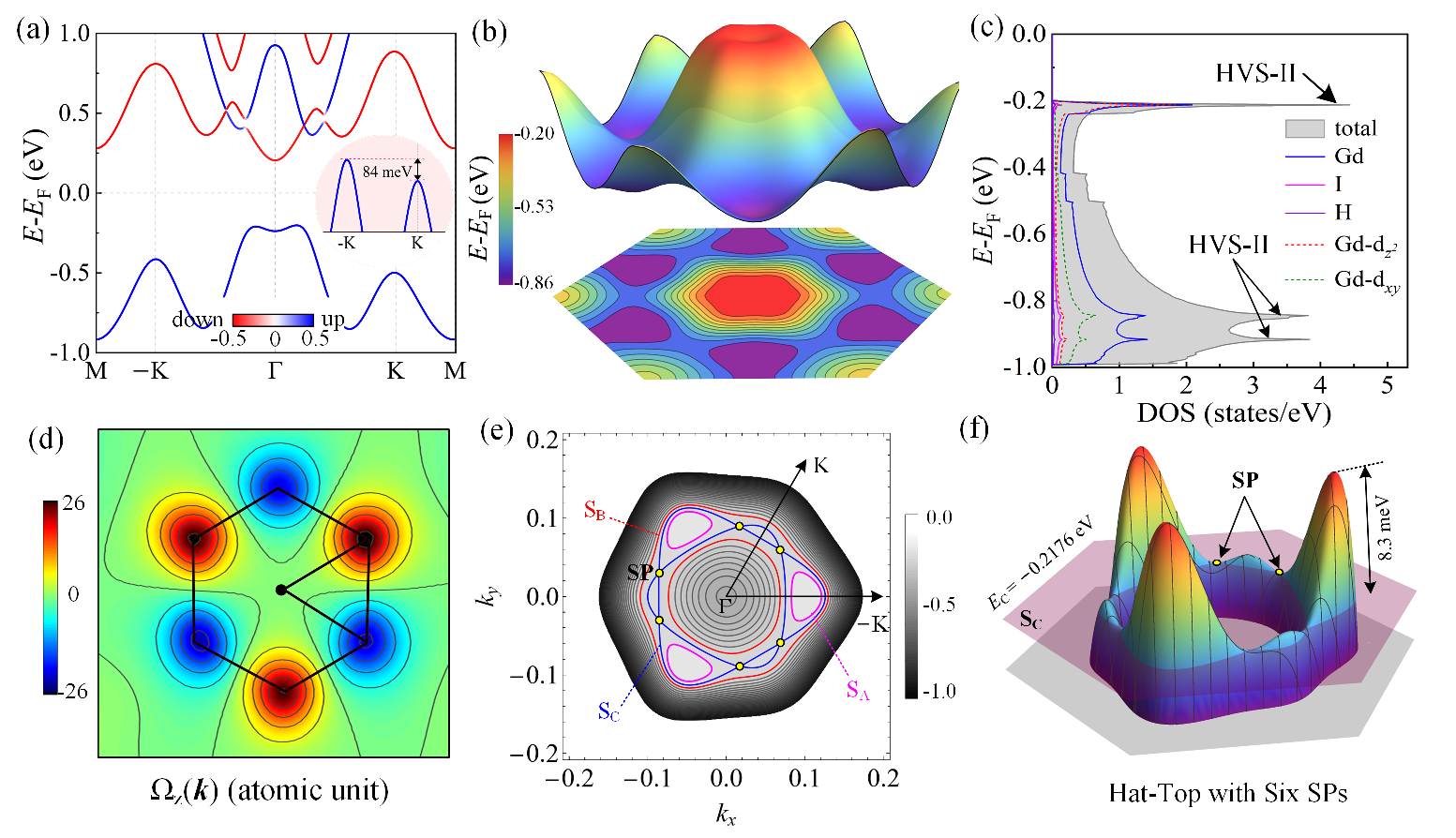}
		\caption{\label{fig:3}(a) The color map of the band structure for GdIH ML with the projection of spin operator $\hat{s}_z$   under considering SOC. (b) The 3D surface \textit{E}($\textit{k}_{x}$, $\textit{k}_{y}$) for the valence band edge in the whole 2D first Brillouin zone. Below is the corresponding 2D contour map. (c) DOS, and atomic and orbital projected DOS in the energy window of [$-1$, 0] eV. (d) The contour plot for the Berry curvatures in the BZ. (e) The energy contours of the fitted Eq.1 on the $\textit{k}_{x}$-$\textit{k}_{y}$ plane for the top of WF-hat, (f) The Zoom-in lens for the 3D hat-top with six SPs.}
	\end{figure*}
	
	\hspace{1.0em}Fig. \ref{fig:3}(a) presents the color map for the spin-polarized electronic band structure of GdIH ML with the projection of the spin operator $\hat{s}_z$ under considering SOC. It shows that GdIH ML features as a bipolar magnetic semiconductor \cite{RN18} with an indirect bandgap of 0.41 eV, in which the valence band edge (VBE) and conductive band edge (CBE) are fully spin-polarized with opposite spin directions. This is highly conducive to the tunability of spin polarization by applying the accessible gate voltage, giving rise to the electrical controlling of magnetism \cite{RN18}. Focusing on the spin-up valence band edge, one can see that it holds the typical features of the proposed WF-hat: (i) a complete WF-hat band appears in the whole first Brillouin zone, as is shown in Fig. 3(b); (ii) the 3D band can be well fitted by using the dispersion model Eq.1 with $m = 7$ (see the fitted parameters in Table S1); (iii) three VHS-II [Fig. \ref{fig:3}(c)] are formed in the corresponding DOS like that in the model case [Fig. \ref{fig:1}h]. These results have also been confirmed by the more accurate HSE06 method (see the bands in Fig. S3a in the Supplemental Material). Additionally, in the following two aspects, we will present some more interesting details about the WF-hat related to the single-spin characteristics.
	
	\textit{Spontaneous valley polarization and high carrier mobility in the hexagonal waved-brim.} Along the waved-brim of the WF-hat, we identify the existence of energy difference [84 meV, see Fig. \ref{fig:3}(a)] between the peaks of the waved-brim at $\rm K$ and $-\text{K}$, indicating that the emergence of spontaneous valley polarization. As is known, in an inversion asymmetrical FM system, two types of spin splitting, i.e., $\Delta_{\text{mag}}$ and $\Delta_{\text{SOC}}$, can usually be induced by the magnetic exchange interaction and SOC effect, respectively \cite{RN71}. For GdIH ML, when the SOC is switched off, the values of $\Delta_{\text{mag}}$ at $\rm K$ and $-\text{K}$ should be equal (see Fig. S3b in the Supplemental Material), namely $\Delta_{\text{mag}}(\rm K) = \Delta_{\text{mag}}(-\text{K})$. However, if the SOC is turned on, the values of $\Delta_{\text{SOC}}$ at $\rm K$ and $-\text{K}$ has opposite sign, i.e., $-\Delta_{\text{SOC}}(\rm K) = \Delta_{\text{SOC}}(-\text{K})$, due to the breaking of inversion symmetry. Therefore, under SOC the spin splitting at $-\text{K}$ and $\rm K$ should be $\Delta_{\text{mag}} - \Delta_{\text{SOC}}$ and $\Delta_{\text{mag}} + \Delta_{\text{SOC}}$, respectively. Then, the valley degeneracy between $-\text{K}$ and $\rm K$ will be eliminated, giving rise to intrinsic ferrovalley ordering  for valley, like the FM state for spin.
	
	To examine the possible applications of the spontaneous valley polarization in WF-hat brim, we compute the Berry curvature $\textbf{$\Omega$}_z(\textbf{\textit{k}})$ [Fig. \ref{fig:3}(d)] by using the Kubo method \cite{RN72} (see Text S2), which is a key parameter for understanding the electronic transport properties. The results reveal that the $\textbf{$\Omega$}_z(\textbf{\textit{k}})$ of GdIH ML shows obvious peaks at both $\rm K$ and $-\text{K}$ valleys but with opposite signs due to the breaking of inversion symmetry. For this, if an external in-plane electric field was applied, the intriguing anomalous valley Hall effect \cite{RN8} can be induced (see the diagram in Fig. S4 in the Supplemental Material), in which additional charge Hall current will appear because the existence of spontaneous valley polarization.
	
		\begin{figure*}[t]
		\includegraphics[width=0.63\linewidth]{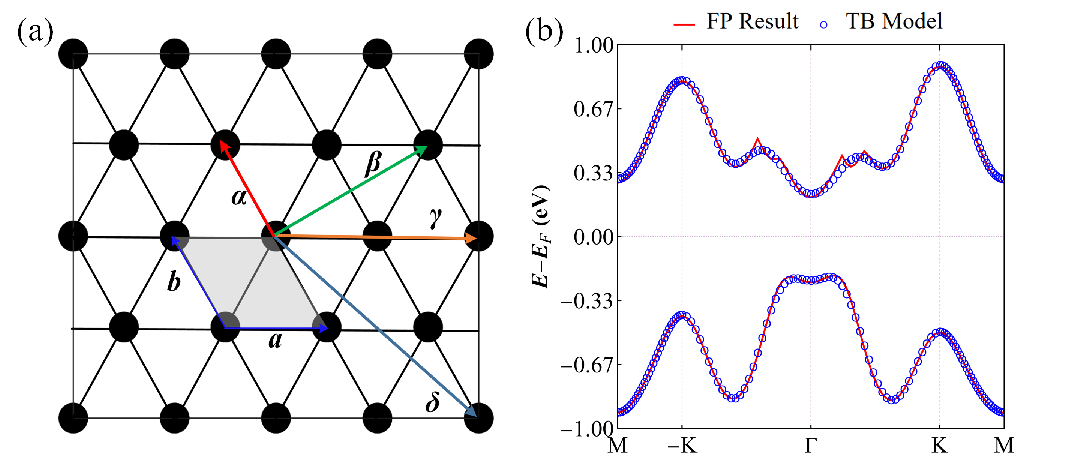}
		\caption{\label{fig:4}(a) The model diagram with nearest neighbor ($\textit{$\alpha$}$), next-neighbor hoping ($\textit{$\beta$}$), $3^{\rm rd}$-neighbor ($\textit{$\gamma$}$) and $4^{\rm th}$-neighbor ($\textit{$\delta$}$) hoping; (b) The fitted TB band structure (blue circle) of GdIH ML together with the FP band one (red line).}
	\end{figure*}
	
	Since the transport properties of carriers are mainly governed by their mobilities, we perform a theoretical prediction for the carrier mobilities of brim $-\text{K}$ valleys in GdIH ML. Under the deformation potential theory \cite{RN73,RN74}, the carrier mobilities of 2D materials can be estimated by the formula: $\mu_{\rm 2D} = e \hbar^3 {\rm C}_{\rm 2D} / k_{\rm B} T m_{\rm e}^* m_{\rm d} E_l^2$. Obviously, there are three key factors, including the effective mass  $m^*$, the deformation potential $E_l$ and elastic modulus ${\rm C_{2D}}$. From the results of them (see Text S3, Table S2 and Figs. S5-S6 in the Supplemental Material), the mobilities of brim holes around $-\text{K}$ are calculated to be $1.1 \times 10^4 \, \text{cm}^2 \, \text{V}^{-1} \text{s}^{-1}$ ($1.4 \times 10^4 \, \text{cm}^2 \, \text{V}^{-1} \text{s}^{-1}$) along $y$($x$)/armchair (zigzag) directions. Owing to the relatively small effective mass stemmed from the massive Dirac dispersion, the carrier mobilities is obviously higher than that of $\rm MoS_{2}$ ML ($0.2 \times 10^3 \, \text{cm}^2 \, \text{V}^{-1} \text{s}^{-1}$ \cite{RN75}), and comparable with that of the prominent black phosphorus ML ($1.0 \sim 2.6 \times 10^4 \, \text{cm}^2 \, \text{V}^{-1} \text{s}^{-1}$ along $y$ direction \cite{RN76}). In this regard, Therefore, it would be possible to obtained the desirable ``high-speed and valley-related'' carrier transport in the future nano-devices based on the hole-doped GdIH ML.

	\textit{Single spin van Hove singularity in the flat top}. Around the central of BZ, the WF-hat top is almost flat, holding almost negligible band dispersion in a very narrow energy window ($\sim$10 meV), far smaller than the energy criterion (25 meV $\sim$ 150 meV) for searching flat-band materials \cite{RN4}. In such window, many-body effects should dominate over the kinetic energy, which may give rise to correlated insulator states, non-Fermi liquid behavior, and the strong-coupling superconductivity \cite{RN77}. When the Fermi level is modulated into the flat top in the vicinity of VBM by applying a gate voltage, the Fermi surfaces would undergo a Lifshitz transition \cite{RN78}, namely a topological geometry transformation: from three isolated half-moon like rings (e.g., $\rm S_{A}$ at energy of $-0.2142$ eV) to two large concentric rings (e.g., $\rm S_{B}$ with energy of $-0.2210$ eV), as illustrated in Fig \ref{fig:3}(e). During such transition, the Fermi level will cross a critical energy of $E_{\rm C} = -0.2176$ eV. On this isosurface ($\rm S_{C}$), six rings with alternating size are interconnected under $\rm C_{\rm 3v}$ point group symmetry.
	
	Interestingly, there are six special touch points, which are correspond to three pairs of saddle points [see SP in Figs. \ref{fig:3}(e) and \ref{fig:3}(f)] located on the two sides of the $\Gamma \rightarrow \rm {K}$ or $\Gamma \rightarrow -\rm {K}$ paths. Around them, the gradient of eigenvalues $E(k_x, k_y)$ along all possible $\textbf{\textit{k}}$ directions is zero, i.e., $\nabla_\textbf{\textit{k}} E = 0$, and the corresponding second-order Hessian matrix ${\rm G} = (\partial^2 E/\partial k_i \partial k_j)$ with $i(j) = x, y$ should be positive semidefinite (i.e., it has both positive and negative eigenvalues) around the six $\textbf{\textit{k}}$ points. Since $g(E) \propto 1/\nabla_k E$ (see Eq. 2), a divergence, namely VHS-II, with $g({\rm E}) \rightarrow \infty$ will emerge in the DOS when $E=E_C$, as is displayed in Fig. \ref{fig:3}(c). Usually, the VHS-II can often give rise to new strongly correlated phases \cite{RN45}, such as superconductivity \cite{RN46} and density waves \cite{RN47}. In particular, the VHS-II of GdIH ML is completely spin-polarized, hence it should be promising for obtaining the desirable ferromagnetic superconductors \cite{RN79,RN80}.
	
	\subsection{A Simplified Two Band Hamiltonian Model for the WF-Hat Band}
	
	\begin{figure*}[t]
	\includegraphics[width=0.65\linewidth]{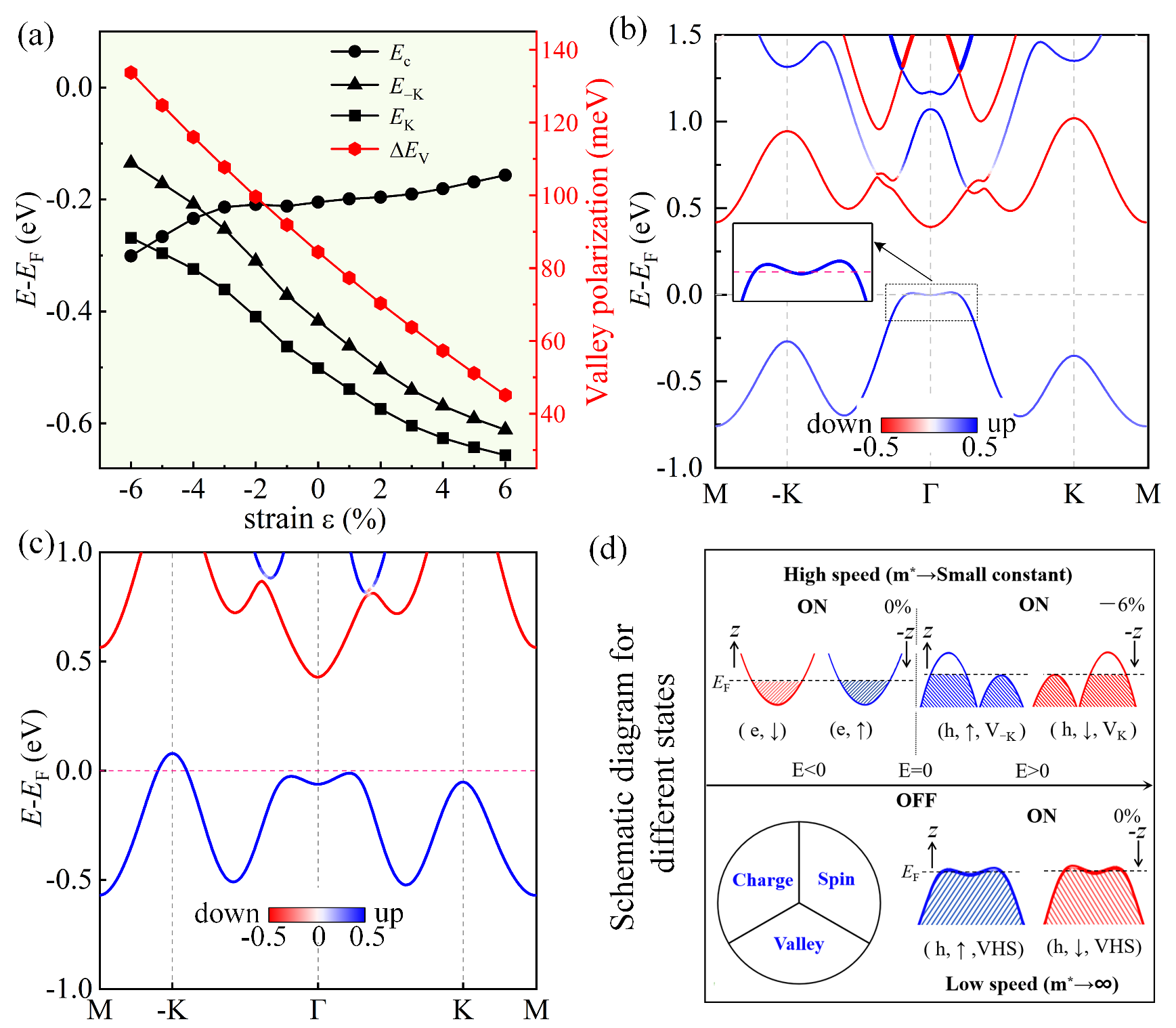}
	\caption{\label{fig:5}(a) The energies for the maximum of the crown of the waved-brim hat band ($E_{\rm c}$), K valley ($E_{\rm K}$) and −K ($E_{-{\rm K}}$) valley, and the values of valley polarization ($\Delta E_{\rm V}$) as a function of external strains ranging from $-$6\% \text{ to } 6\%.(b) Schematic diagram for different states. The color map of the band structures with the projection of spin operator $\hat{s}_z$ under considering SOC at different carrier doping concentrations of (b) 0.1 electrons and (c) 0.1 holes per primitive cell (7.97 $\times 10^{3}$ $\text{cm}^{-2}$).}
    \end{figure*}

	\hspace{1.0em}Considering the FM order, the magnetic space group (MSG) of GdIH ML should be the ``black-white'' group $P3m'1$ (156.51), belong to the type-III MSG without antitranslation symmetry. Under the corresponding symmetry operaties, the localized anionic electron (from 5$d^1$ of Gd atoms) with an equivalent Wycoff position $(0, 0, {\rm z}|0, 0, {\rm m}_{z})$ should obey ${\rm C}_{\rm 3v}$ point group symmetry. For this, the $d$ orbitals would split into three categories: ${\rm A}_1(d_{z^2})$, ${\rm E}(d_{xz}, d_{yz})$, and ${\rm E}(d_{xy}, d_{x^2-y^2})$. Our first-principles fat-band analysis (see Fig. S7 in the Supplemental Material) reveals that the Bloch states of GdIH ML near the Fermi level are mainly contributed by $d_{xy}$, $d_{x^2-y^2}$, and $d_{z^2}$. From the symmetric matching principles based on the molecular orbital theory \cite{RN81}, one can deduce that the vertical mirror symmetries $\hat{\sigma}_v$ allow the orbital hybridization between the ${\rm A}_1$ and the two ${\rm E}$ categories. Therefore, it is reasonable to take $\left| \varphi_1 \right\rangle
	= d_{xy}$ and $\left| \varphi_2 \right\rangle
	= d_{x^2-y^2}$ as the bases for a two-dimensional wave functional space. Then a simplified two-bands effective Hamiltonian can be constructed by considering $4^{\rm th}$ neighbor $d$-$d$ hopping [(Fig. \ref{fig:4}(a)]:
	
	\begin{equation}\label{func_3}
		{\rm H}^{\rm 4th}(\textbf{\textit{k}}) = \begin{pmatrix} {\rm H}_{11} & {\rm H}_{12} \\ {\rm H}_{12}^* & {\rm H}_{22} \end{pmatrix}
	\end{equation}
	
	Using the tight-binding (TB) method \cite{RN82,RN83}, the matrix elements of ${\rm H}^{\rm 4th}(\textbf{\textit{k}})$ can be obtained from ${\rm H}_{\mu \mu'}(\textbf{\textit{k}}) = \sum_R e^{i \textbf{\textit{k}} \cdot \textbf{\textit{R}}} E_{\mu \mu'}(\textbf{\textit{R}})$. Here, $\textbf{\textit{R}}$ is the lattice vector, and the hopping integral $E_{\mu \mu'}(\textbf{\textit{R}})$ can be evaluated by:
	
	\begin{equation}\label{func_4}
		E_{\mu \mu'}(\textbf{\textit{R}}) = \langle \varphi_{\mu}(\textbf{\textit{r}}) | {\rm H}^{\rm 4th}(\textbf{\textit{k}}) | \varphi_{\mu'}(\textbf{\textit{r}} - \textbf{\textit{R}}) \rangle
	\end{equation}For simplicity, the obtained ${\rm H}_{\mu \mu'}(\textbf{\textit{k}})$ are expressed as:
	
	\begin{equation}\label{func_5}
		{\rm H}_{11} = \sum_{n,m} {\rm A}_{n,m} \cos(n k_x + m k_y) + {\rm B}_{n,m} \sin(n k_x + m k_y)
	\end{equation}
	
	\begin{equation}\label{func_6}
		{\rm H}_{22} = \sum_{n,m} {\rm C}_{n,m} \cos(n k_x + m k_y) + {\rm D}_{n,m} \sin(n k_x + m k_y)
	\end{equation}
	
	\begin{equation}\label{func_7}
		{\rm H}_{12} = \sum_{n,m} {\rm E}_{n,m} \cos(n k_x + m k_y) + {\rm F}_{n,m} \sin(n k_x + m k_y)
	\end{equation} in which the coefficients ${\rm A}_{n,m}$, ${\rm B}_{n,m}$, ${\rm C}_{n,m}$, ${\rm D}_{n,m}$, ${\rm E}_{n,m}$, ${\rm F}_{n,m}$ are all polynomials of different hopping parameters, including $\varepsilon_j$, $\alpha_j$, $\beta_j$, $\gamma_j$, and $\delta_j$, which respectively represent the on-site energies, nearest neighbor, next-neighbor hopping, $3^{\rm rd}$-neighbor and $4^{\rm th}$-neighbor hopping parameters (see Table S3). On basis of the TB model, then we solve the eigenvalue function $E(\textbf{\textit{k}})$. Although the list of parameters is relatively complex, they can be well fitted using Quasi Newton method \cite{RN84}. The fitted 26 parameters for $E(\textbf{\textit{k}})$ and ${\rm H}^{\rm 4th}(\textbf{\textit{k}})$ are listed in Table S4. The corresponding two-edge TB bands are displayed in Fig. \ref{fig:4}(b). One can see that the two TB bands agree well with the corresponding first-principles (FP) results. Therefore, the proposed WF-hat can be well described by the effective Hamiltonian model Eq.3, which should be useful in the future study about the transport properties, optical properties, or many-body physics of WF-hat edge band.
		
	\subsection{The adjustability of different electronic states in GdIH ML}
	\hspace{1.0em}For practical device applications, the switchability and adjustability of the intriguing electronic states of a new 2D material are highly desirable. In experiments, there are two readily achievable methods for engineering the electronic properties, including (i) mechanical strains which can be induced by stretching or bending of a flexible substrate \cite{RN85}, and (ii) electrolyte gate which can generate high doping level of electron or hole carriers up to $10^{15}$ cm$^{-2}$ in 2D materials \cite{RN86,RN87}. To explore the impact of external strains on the WF-hat of GdIH ML, we first recalculate its band structures (Fig. S8 in the Supplemental Material) under different strains ranging from $-$6\% to 6\%, which are accessible in experiments. From Fig. \ref{fig:5}(a), one can see that the valley polarizations ($\Delta E_{\rm V}$) between K and -K valleys in the brim almost linearly change with respect to the applied strains, satisfying the expression: $\Delta E_{\rm V} = -7.58 \varepsilon + 84$ (meV).This stems from the similar changing trend of energies at K ($E_{\rm K}$) and $-{\rm K}$ ($E_{-{\rm K}}$) valleys. Interestingly, when the compressive strain is larger than 4\%, the maximum of the hat-top in the $\Gamma \rightarrow {\rm K}$ path is smaller than $E_{\rm K}$, and then smaller than $E_{-{\rm K}}$ under $-$6\% strain. This means that the proposed WF-hat undergoes a noticeable distortion with hat-top sinking and hat-brim rising. Consequently, there forms a clean energy window ($\sim$ 135 meV) occupied only by the $-$K valley band, which provides the possibility to obtain fully valley polarized high-speed holes.
	
	Combined with the mechanical strains, we further examine the possible conductive ON states by applying carrier doping. For unstrained GdIH ML, the flat-top can be well preserved under a modest hole doping ($\sim$ $10^{13}$ cm$^{-2}$), and the Fermi level can be precisely shifted in the hat-top energy window [$\sim$ 10 meV, see Fig. \ref{fig:5}(b)]. If the dispersion is ignored, the $m^*$ should approach to the infinity, and the corresponding mobility of heavy Fermions should be zero, i.e., ultra-low speed. Moreover, the narrow conductive window provides the opportunity to get the Lifshitz transition and single-spin metallic VHS-II state (\textit{h}, $\uparrow$, VHS) as discussed above.

	Under the strain of $-6\%$, hole doping can move the Fermi level immerging into the $-$K valley, forming a single-spin and -valley metallic state $(h, \uparrow, {\rm V}_{-{\rm K}})$, as shown in Fig. \ref{fig:5}(c). From the point of view of carrier mobilities, the hat related $(h, \uparrow, {\rm VHS})$ and $(h, \uparrow, {\rm V}_{-{\rm K}})$ are two extreme opposite states which can be viewed as a ``bipolar'' feature (ultra-low-speed vs. ultra-high-speed), corresponding to heavy Fermions and massive Dirac Fermions, respectively.It should be emphasized that, apart from the states on the hat band, there has another high-speed conductive ON state $(e, \downarrow)$ with a large intrinsic carrier mobility of $1.4 \times 10^4 \, \text{cm}^2 \, \text{V}^{-1} \text{s}^{-1}$ along the $\textit{y}$ direction (see Table S2), which can be induced by electron doping (see Fig. S9 in the Supplemental Material). In contrary to $(h, \uparrow, {\rm V}_{-{\rm K}})$, such state has opposite spin direction due to the bipolar magnetic feature ($\uparrow$ vs. $\downarrow$). In this regard, GdIH ML can be viewed as a rare \textit{double bipolar} semiconductor, being promising for the future multifunctional nanodevices. Fig. \ref{fig:5}(d) presents a schematic diagram for the transition between the mentioned three states under strains and carrier doping. Notably, if the magnetization direction was tuned to $-z$ axis by an external magnetic field, both spin and valley of the three identified states will reverse, changing to $(h, \downarrow, \text{VHS})$, $(h, \downarrow, {\rm V}_{\rm K})$, and $(e, \uparrow)$.

	\section{CONCLUSIONS}
	\hspace{1.0em}In summary, by proposing a general band dispersion model for different hat-like bands, we identify a novel single-spin WF-hat band, which holds a flat top and a waved brim with six valleys. Via first-principles calculations, we demonstrate that a new Janus FM electrene 2D material, GdIH ML, can harbor such special hat state. The interesting results can be summarized in the following points: (1) it is XY magnet with a large MAE and robust FM ground state near ambient temperature; (2) owing to magnetic exchange interaction, its WF-hat is fully spin-polarized, giving rise to single-spin VHS-II on the top and sizable spontaneous valley polarization along the brim; (3) based on group theory analysis, a simplified two-band effective Hamiltonian is constructed, which can well describe the whole WF-hat band; (4) under strains and carrier doping, different conductive ON states, including ($\textit{h}$, $\uparrow$, $\text{VHS}$), ($\textit{h}$, $\uparrow$, ${\rm V}_{-{\rm K}}$), ($\textit{e}$, $\downarrow$), ($\textit{h}$, $\downarrow$, $\text{VHS}$), ($\textit{h}$, $\downarrow$, ${\rm V}_{-{\rm K}}$) and ($\textit{e}$, $\uparrow$) can be obtained, which is promising for multifunctional nanodevices applications; (5) remarkably, the special WF-hat makes GdIH ML become a rare double bipolar semiconductor. 
	
	\textbf{NOTE:} Our results in this work may stimulate further experimental research interest because the synthesized bulk ${\rm GdI}_{2}$ has a quite small exfoliation energy ($\sim0.24 \, \text{J/m}^2$) \cite{RN31} and the similar Janus structure MoSH ML has been synthesized by using gentle $\text{H}_2$ plasma treatment \cite{RN51}. Besides, the further high-throughput searching for other ideal 2D materials with such intriguing hat state in different lattice symmetry is very important and quite urgent. In our following works, this will be an interesting project.
	
	\section*{ACKNOWLEDGMENTS}
    \hspace{1.0em}The authors thank X. Huang for the useful discussions on the construction of Hamiltonian model. This work is supported by Natural Science Foundation of Inner Mongolia Autonomous Region (2021JQ-001), the National Natural Science Foundation of China (12064030, 12264033, 11964023), and the 2020 Institutional Support Program for Youth Science and Technology Talents in Inner Mongolia Autonomous Region (NJYT-20-B02). 
    \section*{AUTHOR CONTRIBUTION}
    \hspace{1.0em}Z L designed the research. N J, Z Y and J C contributed equally to this work. Z Lv, Y S, T S and X C performed all the first-principles calculations supervised by Z L and N J, Z L prepared the manuscript with notable inputs from all authors.
	~\\
	\bibliography{C6-manuscript}
	
	\end{document}